\documentclass[]{spie}
\usepackage{graphicx,amsmath,amssymb,xspace}
\usepackage{cite}

\newcommand{\mum}{\ensuremath{~\mu\mathrm{m}}}
\newcommand{\val}[2]{\ensuremath{#1~\mathrm{#2} \xspace}}

\newcommand{\fratio}{$f$-ratio\xspace}
\newcommand{\filtB}{\val{440}{nm}\xspace}
\newcommand{\filtI}{\val{790}{nm}\xspace}
\newcommand{\filty}{\val{551}{nm}\xspace}

\title{Impact of surface-polish on the angular and wavelength dependence of fiber focal ratio degradation}
\author{Arthur D. Eigenbrot, Matthew A. Bershady, Corey M. Wood
\skiplinehalf
University of Wisconsin, 475 N. Charter St, Madison WI, 53706, USA}
\authorinfo{Correspondence to eigenbrot@astro.wisc.edu}
\begin{document}
\maketitle

\begin{abstract}

We present measurements of how multimode fiber focal-ratio degradation
(FRD) and throughput vary with levels of fiber surface polish from 60
to 0.5 micron grit. Measurements used full-beam and laser injection
methods at wavelengths between 0.4 and 0.8 microns on 17 meter lengths
of Polymicro FBP 300 and 400\mum\ core fiber. Full-beam injection
probed input focal-ratios between $f$/3 and $f$/13.5, while laser
injection allowed us to isolate FRD at discrete injection angles up to
17 degrees ($f$/1.6 marginal ray). We find (1) FRD effects decrease as
grit size decreases, with the largest gains in beam quality occurring
at grit sizes above 5\mum; (2) total throughput increases as grit
size decreases, reaching 90\% at \filtI with the finest polishing
levels; (3) total throughput is higher at redder wavelengths for
coarser polishing grit, indicating surface-scattering as the primary
source of loss. We also quantify the angular dependence of FRD as a
function of polishing level. Our results indicate that a commonly
adopted micro-bending model for FRD is a poor descriptor of the
observed phenomenon.

\end{abstract}


\section{Introduction}

Multimode optical fibers provide the most cost-effective coupling between
telescopes and spectrographs that allow spectrographs to be placed in
stable environments. However, these fiber optics contribute to light
loss from attenuation within the fiber material and surface-scattering
of their ends, and increase entropy in the optical beam. The latter is
referred to as focal ratio degradation (FRD), whereby light injected
into a fiber at a particular \fratio emerges at a faster (smaller)
\fratio. Ever since the first efforts to characterize FRD in
astronomical applications\cite{Angel77} astronomers have attempted
to understand its cause(s) in the hope to lessen its
effects\cite{Carrasco,Oliveira}. Microbends have historically been a
favored culprit\cite{Carrasco,Gloge72}, but recently it has been
suggested\cite{Haynes11, Avila98} that scattering caused by
surface-roughness on the fiber face contributes significantly to
FRD.

We discuss the results of two experiments designed to measure how the
amount of FRD depends on surface roughness, wavelength, and input
angle. The experiments described here use both full-beam and laser
injection methods\cite{Carrasco} standard for FRD tests in
astronomical applications. The full-beam method is useful for
characterizing how a fiber would perform when fed by a telescope,
and provides a straightforward way to compute practical metrics useful
for designing spectroscopic instruments. The laser injection method
allows light to be injected into the fiber at discrete input angles
(compared to a filled ray-bundle cone). This angular dependence of
scattering is a particularly sensitive diagnostic of the physical
mechanisms responsible for FRD.

The method and results of our experiment of FRD dependence on surface
roughness are presented in \S \ref{sec:polish}. Results for the
wavelength dependence of FRD are reported in \S
\ref{sec:wavelength}. The dependence of scattering on input angle is
reported in \S \ref{sec:angle}, and the implications of our work are
discussed in \S \ref{sec:summary}.

\section{Grit Size and Wavelength Dependence}
\label{sec:polish}
\subsection{Polishing Method}
\begin{figure}[ht]
    \includegraphics[width=1.0\textwidth]{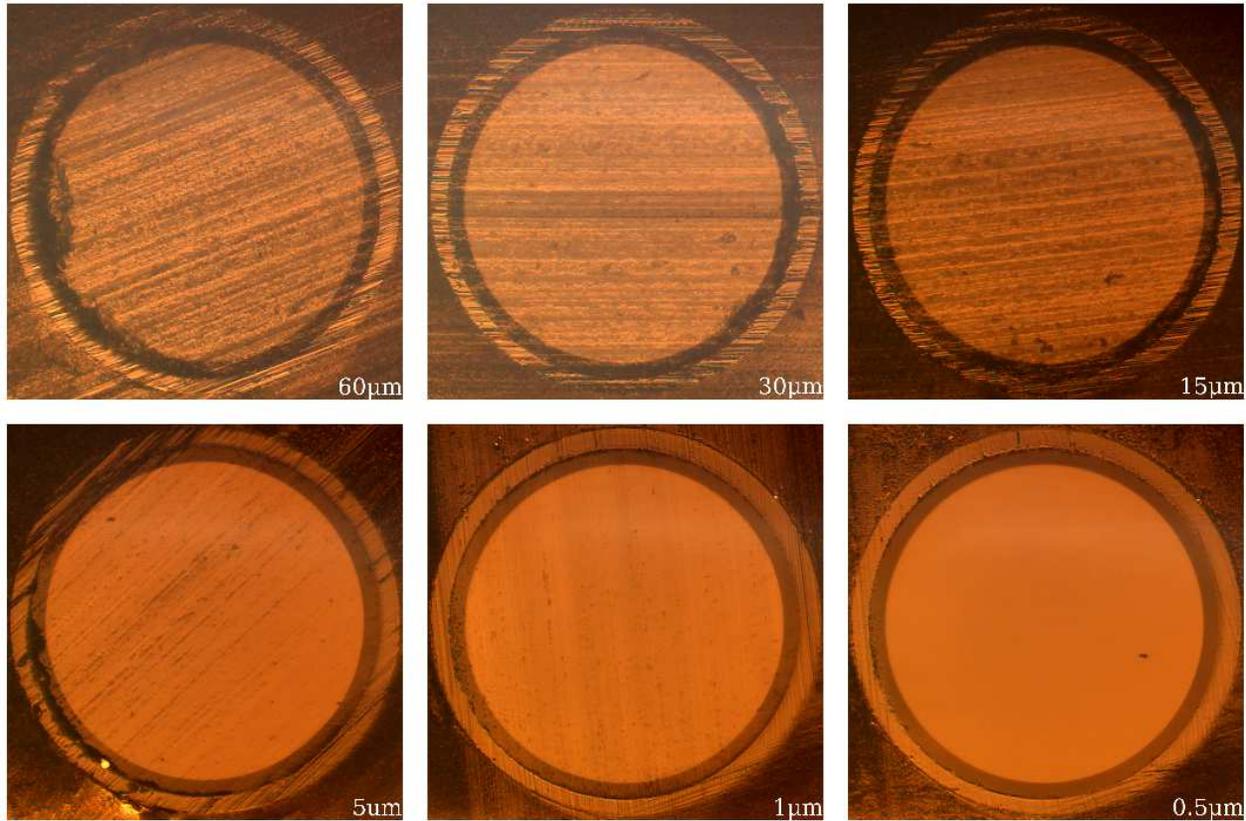}
    \caption{Images of the ``test'' surface of the fiber cable showing
      the appearance of the fiber face at each of the six stages of
      polishing. The number in the bottom right corner of each panel
      denotes the size of grit used. The fiber core diameter is
      300$\mu$m.\label{fig:mosaic}}
\end{figure}

Our tests used Polymicro Technologies stepped-index broadband optical
fibers, cut to $\sim$17 meters in length.  FBP300330370 has
core:clad:buffer diameters of 300:330:370\mum, respectively; a second
length of FBP400440480 has 400:440:480\mum\ diameters.  We mounted
each fiber end into 0.25 inch cylindrical brass ferrules using Norland
Optical Adhesive 61 ultraviolet-curing epoxy.

The experiment was designed to maintain one end of each fiber cable as
a ``control'' surface, well-polished (0.5\mum) at the beginning of
testing, and to use the other end of the fiber cable as the ``test''
surface, initially polished using a coarse grit and then re-polished
with successively finer grit sizes after each measurement.  For
creating the progression of decreasing surface roughness we performed
consecutive polishing steps on the test surface using silicon carbide
lapping disks of 60, 30, 15, 5, and 1\mum\ grit as well as 0.5
\mum\ grit aluminum oxide disks.

The fiber cable was polished using an Ultra Tec Manufacturing, Inc.,
ULTRAPOL 1200-series lapping machine.  We considered a fiber surface
to be adequately polished when, under visual inspection, surface
grooves appeared to be relatively even across the face with very few
surface features larger than the grit size used to polish the surface.
We imaged each end of the fiber using a Newport Corporation F-ML1
fiber inspection microscope for magnification and a Motic Corporation
Moticam 2300 CMOS detector.  We took images after each polishing step
in order to confirm that the test surface was adequately polished and
that the control surface remained undamaged.  We also captured images
after each FRD measurement step in order to determine if the fiber
faces had been damaged during handling, as any potential damage could
affect the test results.  A mosaic of images of the test surface, as
seen after the FRD measurement at each grit size, is seen in
Figure~\ref{fig:mosaic}.

\subsection{Data Collection}
\label{sec:direct}
The experimental apparatus is a modified version of the far-field
differential beam comparator using a double re-imaging system
described in Crause et al. (2008)\cite{Crause_08}, which is
based of the FRD test apparatus used to characterize the fibers of
SPARSPAK\cite{Mab_04}. The final collimating lens (L3 in Crause et
al. (2008)\cite{Crause_08}) was replaced with a Canon $f$/1.2 camera
lens to eliminate ghost images caused by very fast output beams. Stray
light was further reduced by covering everything downstream of the
focus plane with a black photographer's cloth. The fiber input stage
was replaced with a highly stable, purpose-built three-axis
translation stage that ensures the fiber face is telecentric to within
$0.01^{\circ}$. Finally, a filter magazine was added to the pinhole
assembly to facilitate the rapid changing of filters.

\begin{figure}[ht]
\centering
\includegraphics[width=0.25\textwidth]{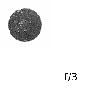}\includegraphics[width=0.25\textwidth]{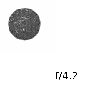}\includegraphics[width=0.25\textwidth]{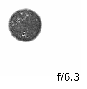}\includegraphics[width=0.25\textwidth]{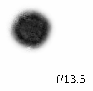}\\
\includegraphics[width=0.25\textwidth]{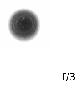}\includegraphics[width=0.25\textwidth]{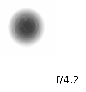}\includegraphics[width=0.25\textwidth]{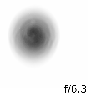}\includegraphics[width=0.25\textwidth]{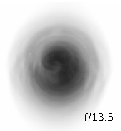}
\caption{\label{fig:dfim}Far-field images of the direct-beam (top) and
  fiber-beam (bottom) outputs for all $f$-ratios tested. Data shown is
  at \filty for the FBP300330370 fiber, 17m in length. Direct and fiber beam images for a given \fratio have identical spatial scales, and are adjusted such that the direct beam images are the same apparent size for all $f$-ratios.}
\end{figure}

Basic operation involves recording data from two imaging modes; the
far-field fiber output, and the far-field image of the direct
beam. The level of FRD is then computed by comparing the two modes.
Examples of these images are shown in Figure \ref{fig:dfim},
illustrating the increasing impact of FRD at slower beam-speeds. Data
were taken in three filters: Johnson I and B and Stromgren \emph{y},
with central wavelengths of \filtI, \filtB, \filty, and widths of
$\Delta\lambda/\lambda=$ 0.22, 0.19, 0.045, respectively; four
$f$-ratios ($f$/3, $f$/4.2, $f$/6.3, and $f$/13.5); and five polish
levels (60, 30, 15, 5, and 1\mum). The range in $f$-ratios was chosen
so that results are applicable to a wide range of telescope
designs. The Wisconsin Indiana Yale NOAO (WIYN) 3.5m telescope has a
$f$/6.3 Nasmyth port; the Sloan Digital Sky Survey Telescope fibers
are fed at $f$/5; and the South African Large Telescope (SALT) feeds its
prime focus instrument package at $f$/4.2 (similar to the Hobby Eberly
Telescope).

Throughput measurements require very precise knowledge of the
intrinsic lamp output (the light input to the system). Experiments in
lab showed that, while the stochastic variations of the lamp are
negligible, there is a secular drift towards lower lamp
output. Subsequent to the experiment reported here a photo-diode
monitoring system has been implemented to correct for this trend. The
data presented here accounted for this trend by: 1) taking a set of
images for one filter with the fiber in place, 2) taking a set of
direct beam images in the same filter, 3) repeating this alternating
scheme until three groups of fiber images, separated by two groups of
direct beam images were taken. This alternating fiber-direct-fiber
scheme is used to remove stochastic lamp variability.

Images were cleaned of detector artifacts and combined to produce five
images for each filter at each polish level: three fiber images and
two direct beam images. Lamp variations were removed by using the
three fiber images as data points to determine a lamp normalization
(relative to the first fiber image) as a function of time. This
function was then used to find normalizations for the direct beam
images. The lamp was found to have no significant variation (10\% of the shot noise)
 within the time required to take one set of data in a particular filter.

\subsection{Analysis}
A custom data reduction pipeline was created to consistently and
efficiently analyze the large volume of data associated with the
multiple filters and polish levels. Analysis consisted of measuring
the total light contained within annuli of constant width and
increasing radius centered on the center of the beam. This information
was then used to construct a curve of growth that shows the fractional
encircled energy (EE) as a function of radius.

In addition to FRD, there are aberrations inherent in our test
apparatus that will affect both the direct beam and fiber beam in the
same way. To remove the effects of these aberrations the difference
between the direct beam and a theoretical ideal beam (no aberrations)
is computed and applied to both the fiber and direct beams; Crause et
al. (2008)\cite{Crause_08} provide a complete description of the
method used. Once the corrections have been applied the only
differences between the fiber and direct beams are caused by FRD.

\subsection{Results}
\label{sec:results}

\begin{figure}[ht]
\begin{center}
\includegraphics[width=\textwidth, trim=0 2.6in 0 0, clip=true]{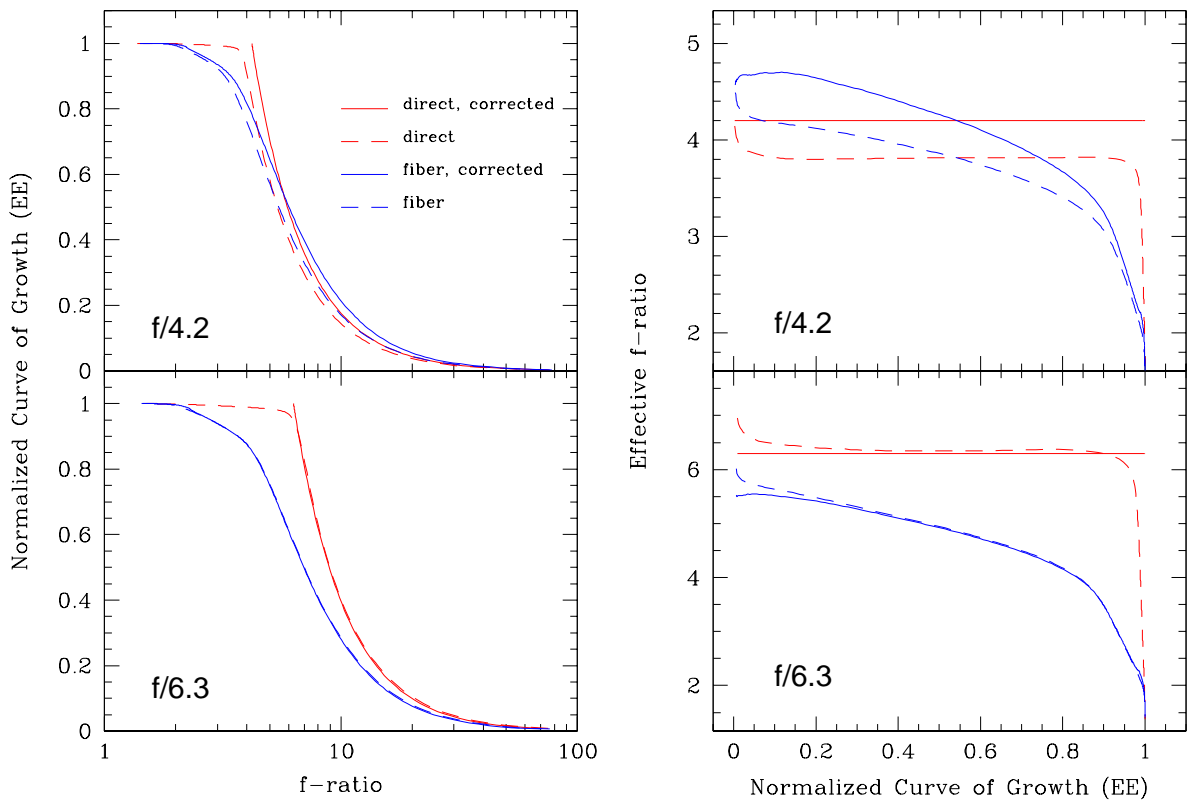}
\caption{\label{fig:basicFRD} FRD effects at \filty, $f$/4.2 and
  $f$/6.3, and with the output face polished to 30\mum. The left
  panels show enclosed energy (EE) as a function of \fratio. The right
  panels show the effective \fratio of a beam that captures a certain
  percentage of the total light (EE). Data is for the FBP300330370
  fiber, 17m in length. A color version of this plot is available online.}
\end{center}
\end{figure}

Figure \ref{fig:basicFRD} shows a characteristic set of FRD analysis
plots at a 30\mum\ end-polish at $f$/4.2 and $f$/6.3.  (See Figure
\ref{fig:grit} for more grit-sizes and \S\ref{sec:gritwave} for
discussion).  The left panels, which consist of normalized curves of
growth (COG) for both the direct and fiber beams, show how FRD
scatters light out to larger radii. The dashed and solid lines are the
data before and after the correction described above\cite{Crause_08}.

The right panels plot the effective \fratio as a function of
EE. These plots only show information about relative light
(re)distribution; they do not contain any information about total
throughput.  Where the fiber curve intersects the ideal beam tells us
what percentage of the original beam's information is being captured
by a spectrograph with an \fratio equal to that of the optics feeding
the fibers. This plot can also be used to estimate how much faster a
spectrograph would have to be to capture more of the input beam. For
example, from figure \ref{fig:basicFRD} and for fibers polished to 30\mum, an $f$/4.2
spectrograph only captures about 53\% of the light fed into the fibers 
at $f$/4.2. If we wanted to capture 90\% of the input light the
spectrograph optics would need to be $f$/3.2.

\subsubsection{Grit-size and \fratio Dependence}
\label{sec:gritwave}
\begin{figure}[htp]
  \centering
  \includegraphics[width=\textwidth]{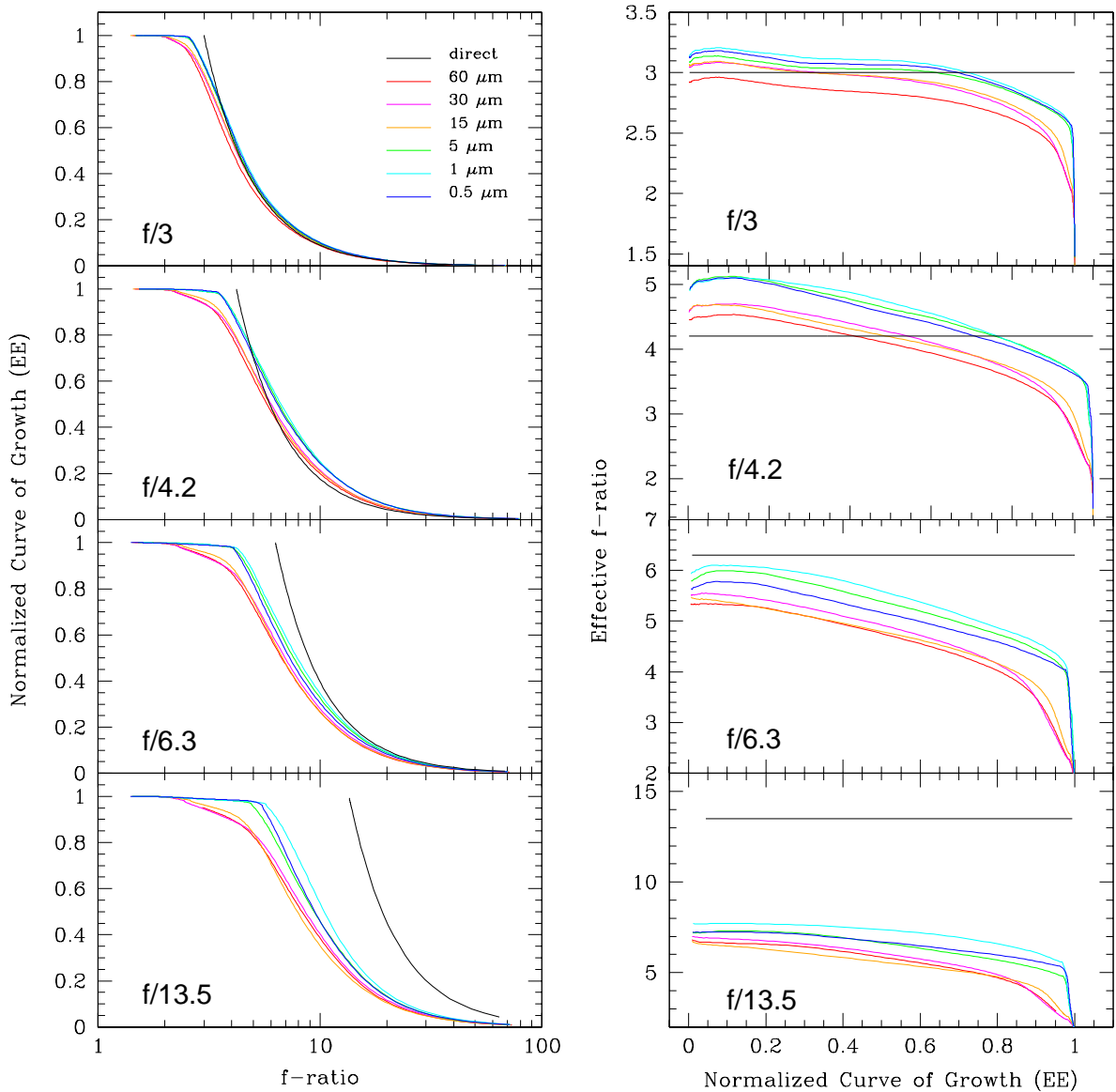}
  \caption{\label{fig:grit}Dependence of FRD on polish grit size at
    \filty and all $f$-ratios measured for 17m of FBP300330370 fiber. A color version of this plot is available online.}
\end{figure}

The primary result of the experiment can be seen in figure
\ref{fig:grit}, which shows FRD curves at one wavelength (\filty) and at
all of the different end-polish levels and $f$-ratios. As expected,
the effects of FRD improve as grit size decreases, but only to a
point. There is steady improvement in output beam quality between 60
\mum\ and 15\mum, a sharp improvement between 15\mum\ and 5\mum,
and almost no improvement between 5\mum\ and 0.5\mum.

It is also worth noting that for certain combinations of input \fratio
and polish level there is a radius (output \fratio) within which there
is relatively \emph{more} light in the fiber beam than in the direct
beam.  The explanation is straightforward: FRD scatters light from
each input angle into both larger and smaller output angles. If the
width of the scattering profile increases towards smaller angles (see
\S\ref{sec:angle}) then more light is scattered out of these angles
compared to larger angles. However, for a uniform input beam
larger input angles contain more luminosity (because they contain
larger annular areas) and so the amount of light scattered in from large
angles will exceed the amount of light scattered out from small angles
despite the wider scattering profile at small angles. In this case the
fiber beam will have relatively more light at smaller angles than the
direct beam, as seen in the plots for $f$/3 and $f$/4.2. Conversely,
there is some sufficiently large output angle that has
significant scattering contributions from angles where there is no
light in the input beam. At these output angles, the COG drops below
the ideal beam, as observed.

It is well known that FRD changes with changing input $f$-ratio,
increasing with slower beams (Ramsey 1988)\cite{Ramsey88}, as seen in Figure
\ref{fig:grit}.  For beams slower than $f$/6.3 the scattering becomes so
large that the output beam never contains more light than the input
beam for any angle.

\subsubsection{Wavelength Dependence and Total Throughput}
\label{sec:wavelength}
\begin{figure}[ht]
  \centering
  \includegraphics[width=\textwidth, trim=0 2.6in 0 0, clip=true]{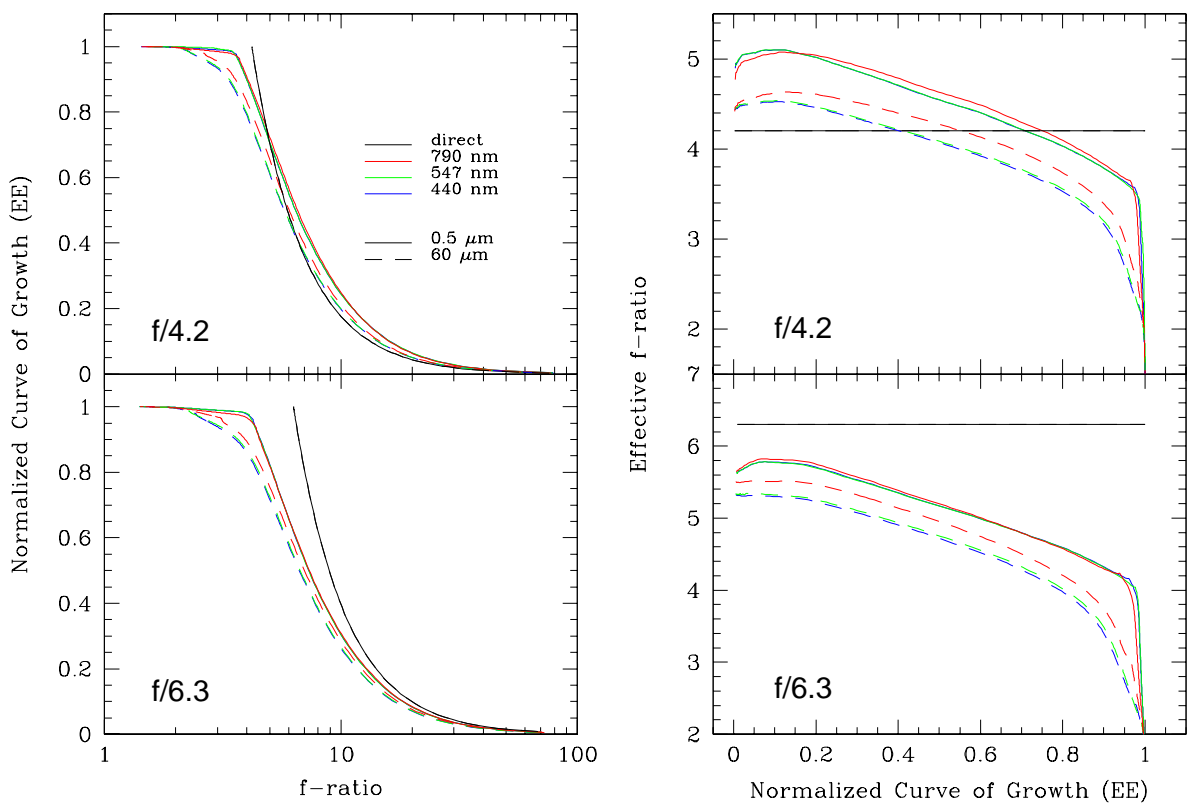}
  \caption{\label{fig:wave} Wavelength dependence of FRD at 0.5 and 60
   \mum\ polish grit at $f$/4.2 and $f$/6.3 for 17m length
    FBP300330370 fiber. A color version of this plot is available online.}
\end{figure}

\begin{figure}[ht]
  \centering
  \includegraphics[width=\textwidth, trim=0 4in 0 0, clip=true]{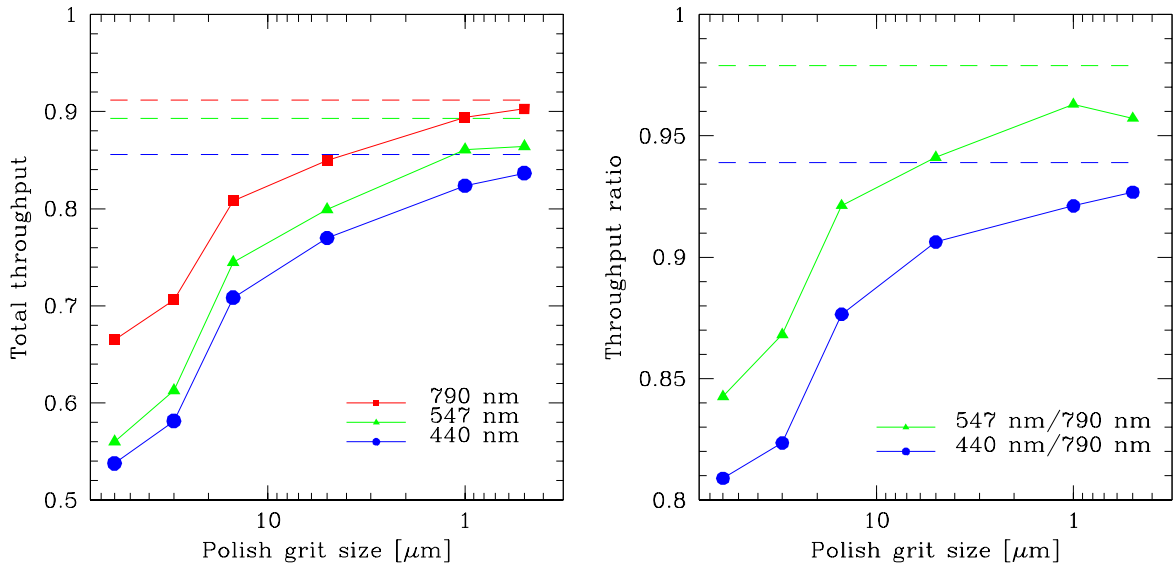}
  \caption{\label{fig:tputwave} Total throughput as a function of
    polish grit for three different wavelengths (left), and the relative
    increase in \filtB and \filty referenced to \filtI (right). Dashed
    lines show the values expected from the product
    specifications. This is measured for a 17m length of FBP400440480
    fiber. A color version of this plot is available online.}
\end{figure}

Previous tests for a wavelength dependence on FRD have been split
between results that suggest there is such a
dependence\cite{Carrasco,Gloge72}, based on what would be predicted by
a micro-bend origin, and results that point to no wavelength
dependence\cite{Mab_04, Schmoll_03}. Figure \ref{fig:wave} shows FRD
curves for all filters at $f$/4.2 and $f$/6.3 and 0.5\mum\ and 60
\mum\ polish levels. The data suggest that at a fine polish level (0.5
\mum) the amount of scattering caused by FRD does {\it not} depend on
the wavelength of the input light. However, at 60\mum\ we do see some
wavelength dependence to the FRD curves, which  must therefore be
caused by surface scattering (see figure \ref{fig:tputwave}).

We also find that the total throughput of the fiber depends on
fiber polish. Figure \ref{fig:tputwave} shows the
total throughput as a function of grit size for all three wavelengths
and $f$/6.3. The total throughput is defined as the asymptotic (in output
 angle) fiber beam counts referenced to the same measurement of the direct
 beam. From the manufacturer's specifications
we expect the fiber attenuation to depend on wavelength. Polymicro
reports attenuations of 20 dB/km at \filtB, 10 dB/km at \filty, and 5
dB/km at \filtI, and we also expect a 3.43\% light loss from each
air-silica interface (input and output faces). Thus, in the case of an
ideal, \val{17}{m} long fiber we expect a throughput of 85.6\%,
89.3\%, and 91.2\% for \filtB, \filty, and \filtI, respectively. These
values are plotted as horizontal dashed lines in figure
\ref{fig:tputwave}. Any additional losses are likely due to surface
scattering.

We also find a polish dependence on the color of the transmission.  As
seen in the right panel of figure \ref{fig:tputwave}, the throughput
gains are greater at shorter wavelengths as the surface-polish
improves, as would be expected from a surface-scattering phenomenon.

\section{Angle Dependence}
\label{sec:angle}
\subsection{Experiment}
We are also interested in the dependence of FRD on the light input
angle. The direct beam injection method reported in \S
\ref{sec:direct} injects a full cone of light into a fiber, i.e., 
the input beam contains rays incident on the fiber face from normal up
to half of the vertex angle of the cone.  To probe FRD effects at
single, discrete input angles (essentially an annular cone) we use an
experiment similar to the laser injection method of Carrasco and
Parry (1994)\cite{Carrasco, Haynes11}, but modified so that the far
field fiber output is imaged on to an opaque screen rather than through a
translucent screen. This was done to eliminate ring blurring that
occurs when imaging through a translucent object of finite thickness.

We also needed to ensure that the imaging screen was far enough away
from the output fiber face. The output images consist of a ring with a
radius corresponding to the laser input angle and a thickness that
varies depending on the amount of FRD present at that particular input
angle. We expect each ring to have an inherent width equal to the
diameter of the fiber, but this width is in a collimated beam while
any FRD scattering results in a slightly diverging beam. With this in
mind, the far field images were recorded at a far enough distance from
the fiber output face that the FRD smearing width described by
Carrasco and Parry (1994)\cite{Carrasco} and Haynes et
al. (2011)\cite{Haynes11} (calculated to be $\sim 2.6^{\circ}$ for our
fiber) dominated the ring width. At the distance chosen the 300
\mum\ core of the fiber subtends $\sim 0.06^{\circ}$ on the screen.
Unfortunately, our direct-imaging approach requires the fiber output and
camera to be off-axis (due to physical constraints), resulting in
elliptical ring images. Significant effort was expended to model the
geometric distortion caused by this method; our analysis software does
an excellent job of removing the distortion to produce circular rings.

Data were taken at input angles of $\pm17^{\circ}$ ($\sim f$/1.6) in
increments of $0.5^{\circ}$. For simplicity we use the FWHM of the ring profile
as a first-order measure of for the amount of FRD, ignoring here the intricacies of
the profile shape\cite{Haynes11,Carrasco}.

\subsection{Results}

\begin{figure}[ht]
\begin{center}
\includegraphics[width=\textwidth,trim=0 0.3in 0 3.8in,clip=true]{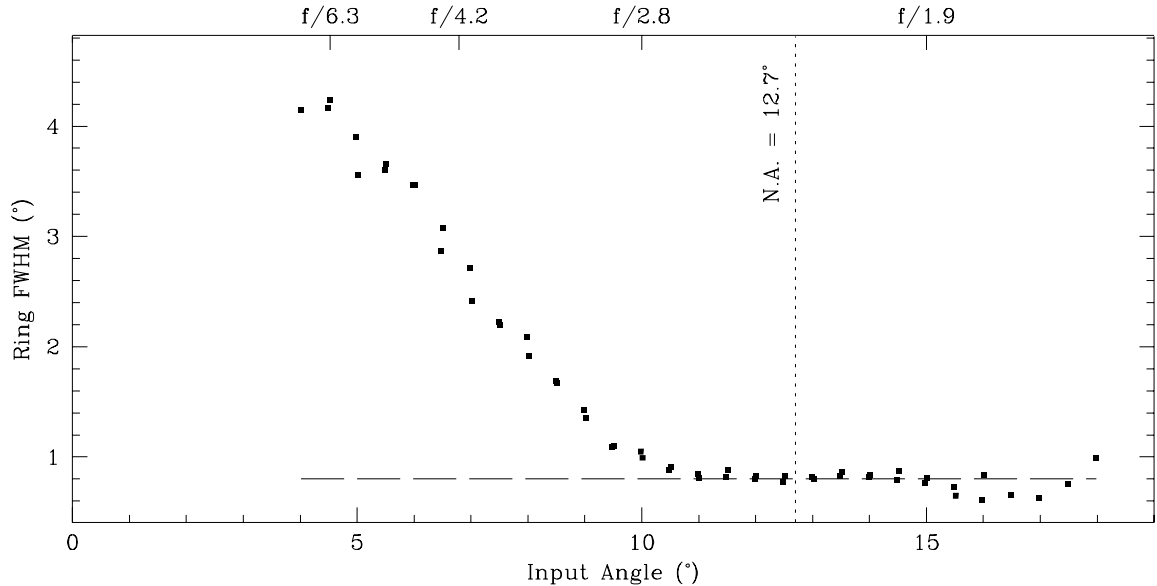}
\caption{\label{fig:angle}
Smearing due to FRD (ring width) as a function on input angle. The vertical dotted line marks the numerical aperture of the fiber specified by the vendor. The horizontal dashed line represents a best fit to the micro-bend model of Carrasco and Parry 1994\cite{Carrasco} over the full angular range of our data adopting $D=\val{1.15\times 10^{-6}}{m^{-1}}$. This value of $D$ is constrained by a best-fit of the micro-bend model to our data in the range of 10 to 18 degrees.}
\end{center}
\end{figure}

Figure \ref{fig:angle} shows how the amount of ``smearing'' caused by
FRD (ring width) varies with different input beam angles. Angles below
$4^{\circ}$ are not plotted; future reports of this experiment will
measure ring widths in the regime where ring-width exceeds ring
radius.

The results in figure \ref{fig:angle} show that FRD effects are very
large at small angles, as expected, and decrease with increasing
angle until leveling out at $\sim 10^{\circ}$, which is within a few
degrees of the numerical aperture (12.7$^{\circ}$). It is possible to
extend the ring measurements to angles above the numerical
aperture. While the throughput decreases in this regime, the FRD
appears to remain fairly constant.

We have compared these trends to predictions of the micro-bend model
presented in Carrasco and Parry (1994)\cite{Carrasco}.  We find a micro-bend parameter of $D=\val{1.15\times 10^{-6}}{m^{-1}}$ provides the best fit to our data (plotted in figure \ref{fig:angle}). This is a lower limit on the value of $D$ for this fiber.
An upper limit on the value of $D$ comes from matching the micro-bend model to our widest ring ($4^{\circ}$), from which we find $D<\val{2.0\times 10^{-5}}{m^{-1}}$. For this range of $D$, our measurements  are in an angular regime where the
micro-bend model predicts a ring width nearly constant in angle. The
fact that our data show a non-constant ring-width indicates that this
micro-bend model is not a good description of the data and therefore not likely the physical origin of FRD.

Our results are also important because they indicate just how strongly
FRD is dominated by light entering the fiber at smaller angles with
respect to the fiber optical axis. In astronomical applications,
fibers are typically fed by on-axis optical systems with central
obstructions from secondary and/or field-corrector optics. From an
entropy stand-point, i.e., information gathering-power per unit cost,
the central rays obscured are the least valuable. Consequently,
wide-field survey-telescopes (e.g., SDSS\cite{York00,Gunn06}), which are fast and have
relatively large central obstructions, are ideally suited to fiber
coupling. Further analysis of data like that in Figure 7 will allow us
to quantify this statement.

\section{Summary}
\label{sec:summary}
Two experiments to 
measure the
amplitude of FRD as a function of wavelength, surface polish, 
and light-injection angle have been carried out and described.
The primary results are:
\begin{enumerate}

\item A component of FRD is attributable to the end-polish
  on fiber surfaces, however this appears to be a second-order effect
  relative to the impact of light-injection angle (beam speed) slower
  than $f$/3. FRD decreases with polishing down to finer grit sizes, but
  not significantly below grit-sizes of 5\mum.

\item Total throughput (light emerging at all angles) also depends on
  end-polish, with a wavelength dependence that indicates the increase
  in throughput is simply a reduction in surface-scattering.  The most
  significant gains occur for polishing that proceeds down to 5 $\mu$m
  grit, although for most astronomical applications at low
  light-levels polishing down to the finest grit is measurably
  advantageous.

\item The amount of FRD does \textbf{not} depend on wavelength, as
  found now in several experiments. This is in contrast to some
  results and the predictions of micro-bend theory for FRD's
  origin\cite{Carrasco}.

\item FRD is dominated by light entering the fiber at smaller angles
  (in our case $<10^{\circ}$), as is well known. Measurements here
  allow us to quantify this statement in detail. The amplitude and
  angular dependence of FRD also do not agree with predictions from
  micro-bend theory.

\end{enumerate}

This work was supported by NSF grants ATI-0804576 and AST-1009471.

\bibliography{SPIEBib}
\bibliographystyle{spiebib}

\end{document}